\documentclass[usenatbib]{mn2e}
\usepackage{graphicx}

\newcommand{\apj}{ApJ}
\newcommand{\apjl}{ApJ}
\newcommand{\apjs}{ApJS}

\newcommand{\aj}{AJ}
\newcommand{\mnras}{MNRAS}
\newcommand{\pasp}{PASP}
\newcommand{\pasj}{PASJ}

\def\eps@scaling{1.0}%
\newcommand\epsscale[1]{\gdef\eps@scaling{#1}}%
\newcommand\plotone[1]{%
 \centering
 \leavevmode
 \includegraphics[width={\eps@scaling\columnwidth}]{#1}%
}%
\newcommand\plottwo[2]{%
 \centering
 \leavevmode
 \columnwidth=1.0\columnwidth
 \includegraphics[width={\eps@scaling\columnwidth}]{#1}%
 \hfil
 \includegraphics[width={\eps@scaling\columnwidth}]{#2}%
}%

\newcommand\acknowledgments[1]{#1}
\newcommand\facility[1]{#1}

\newcommand\ion[2]{#1$\;${\small{#2}}}

\usepackage{url}

\newcommand{\tauGPeff}{\tau_{\rm GP}^{\rm eff}}
\newcommand{\xHI}{\bar{x}_{\rm HI}}
\newcommand{\zfivesix}{z \sim 5\mbox{--}6}
\newcommand{\sigone}{\sigma_{\rm 1 LOS}}
\newcommand{\Nlos}{N_{\rm LOS}}

\newcommand{\avenf}{\xHI}
\newcommand{\lsim}{\la}
\newcommand{\gsim}{\ga}
\newcommand{\lya}{Ly$\alpha$}
\newcommand{\lyb}{Ly$\beta$}
\newcommand{\Msun}{M_\odot}

\newcommand{\HI}{\ion{H}{I}}
\newcommand{\HII}{\ion{H}{II}}
\newcommand{\MgII}{\ion{Mg}{II}}
\newcommand{\OVI}{\ion{O}{VI}}

\begin{document}

\title[Upper limit on the neutral fraction at $\zfivesix$]{The first (nearly) model-independent constraint on the \\
 neutral hydrogen fraction at $\zfivesix$}

\author[I. D. McGreer, A. Mesinger, \& X. Fan]{
Ian D. McGreer,$^{1,}$\thanks{Email: imcgreer@as.arizona.edu}
Andrei Mesinger,$^{2,}$\thanks{Hubble Fellow; Email: mesinger@astro.princeton.edu} and
Xiaohui Fan$^{1}$\\
$^{1}$ Steward Observatory, The University of Arizona, 
                 933 North Cherry Avenue, Tucson, AZ 85721--0065 \\
$^{2}$ Department of Astrophysical Sciences, Princeton University,
                 Princeton, NJ 08544, USA \\
}

\voffset-0.6in

\maketitle

\begin{abstract}
Cosmic reionization is expected to be complex, extended and very inhomogeneous. 
Existing constraints at $z\sim6$ on the volume-averaged neutral hydrogen 
fraction, $\avenf$, are highly model-dependent and controversial.  Constraints 
at $z<6$, suggesting that the Universe is highly ionized, are also 
model-dependent, but more fundamentally are invalid in the context of 
inhomogeneous reionization.  As such, it has recently been pointed out that there
is no conclusive evidence that reionization has completed by $\zfivesix$, a fact 
that has important ramifications on the interpretation of high-redshift 
observations and theoretical models. We present the first direct upper limits on 
$\avenf$ at $\zfivesix$ using the simple and robust statistic of the covering 
fraction of dark pixels in the Ly$\alpha$/$\beta$ forests of high redshift 
quasars. With a sample of 13 Keck ESI spectra we constrain 
$\avenf\lsim0.2$ at $5<z\lsim5.5$, rising to $\avenf\lsim0.8$ at $z\sim6.1$.  
We also find tentative evidence for a break in the redshift evolution of the 
dark covering fraction at $z\sim5.5$. A subsample of two deep spectra provides
a more stringent constraint of $\xHI(z=6.1) \la 0.5$ when combined with 
conservative estimates of cosmic variance. This upper limit is comparable to 
existing results at $z\sim6$ but is more robust.  The results presented here
do not rely on assumptions about the quasar continuum, IGM density, \HII\ 
morphology or ionizing background fields, and thus are a good starting point for 
future interpretation of high redshift observations.
\end{abstract}


\section{Introduction}\label{sec:intro}

The reionization of the Universe by the first generations of astrophysical 
sources is of fundamental importance, offering glimpses into the early Universe 
and insight into many astrophysical processes.  Consequently, much effort has 
gone (and continues to go into) understanding this epoch. Many indirect 
observational probes of reionization have been proposed. Arguably foremost 
among these involve the spectra of $z\sim6$ quasars, first discovered with the 
Sloan Digital Sky Survey \citep[SDSS;][]{Fan+01,Fan+03,Fan+04,Fan+06qsos}.

Quasars have long been useful probes of the IGM across a wide range of
redshifts through Ly$\alpha$ forest studies. The moderate redshift 
($z\lsim3$) Ly$\alpha$ forest is largely transparent, with occasional 
transmission gaps arising from strong absorption systems, e.g., Lyman Limit 
Systems (LLSs) and Damped Ly$\alpha$ Absorbers (DLAs). However, at higher 
redshifts ($z\gsim5$), the \lya\ forest becomes increasingly opaque and quasar 
absorption spectra saturate \citep{Becker+01,Djorgovski+01} due to the strong 
absorption cross-section of the \lya\ transition.  Even trace amounts of 
neutral hydrogen (with neutral fractions as low as 
$x_{\rm HI}\sim10^{-5}$--$10^{-4}$) are then sufficient to render the forest 
optically thick. Furthermore, reionization driven by stellar sources progresses 
in a highly inhomogeneous manner, with neutral patches of the IGM becoming 
increasingly rare. Thus, robust and tight constraints on reionization from the 
\lya\ forest are difficult to obtain and require detailed modeling.  

Modeling reionization is a significant challenge, as simulations need to be 
large enough to statistically capture the biased locations of quasar host halos 
\citep{Lidz+07,AA07, M10} as well as the rare voids that dominate the 
transmission at high redshifts (e.g., \citealt*{Becker+07}).  At the same time, 
simulations must have sufficient resolution to resolve the dominant ionizing 
population expected to reside in atomically cooled, $M\sim10^8\Msun$ halos 
(e.g., \citealt*{MD08,CFG08}), and the small-scale structure in the IGM 
\citep*[e.g.,][]{Lidz+06}.  More approximate techniques have recently been 
developed in order to overcome the challenges posed by such a large range of 
scales, including sub-grid \citep{KG07,McQuinn+07b} and semi-numerical 
\citep[e.g.,][]{MF07} methods.

Perhaps the most prominent metric to have come out of the high-redshift forest 
studies of reionization is the so-called Gunn-Peterson \citep[GP,][]{GP} 
effective optical depth, $\tauGPeff$.  This quantity is defined somewhat 
arbitrarily (for lack of an obvious alternative) by the logarithm of the mean 
flux, $\tauGPeff \equiv - \ln \langle F_{\rm GP} \rangle$, and is dependent on 
the complex structure of the density, ionization and UV background fields 
\citep{Fan+02,SC02}.  Nevertheless, measurements of $\tauGPeff$ have led to two 
important conclusions: (i) we might be witnessing the final stages of 
reionization, as evidenced by the steep increase with redshift in the mean and 
scatter of $\tauGPeff$ at $z\gsim6$; (ii) the Universe is highly ionized at 
$z\lsim6$, as evidenced by the lack of such a rise and scatter.

The first conclusion has been the subject of much recent controversy. 
Specifically, empirical extrapolations with alternative (though somewhat 
arbitrary) models of the density field and continuum fitting can reproduce the 
steep rise in $\tauGPeff$ without having to appeal to reionization 
(\citealt{Becker+07}, see also \citealt{Songaila04}). Additionally, the 
observed scatter in $\tauGPeff$ is consistent with it being driven solely by 
the density field \citep{Lidz+06}.  In fact, it is unlikely that the final 
overlap stages of reionization are accompanied by a steep rise in the UV 
background (or an associated fall in $\tauGPeff$) as was initially believed 
\citep[e.g.,][]{Gnedin00}, since the evolution of the ionizing photon mean free 
path in these end stages is likely regulated by LLSs and not reionization 
itself \citep{FM09}.

The second conclusion derived from $\tauGPeff$ -- that the Universe is highly 
ionized at $z\lsim6$ -- has been far less controversial.  However, similar 
arguments about the ambiguity of the observations at $z\sim6$ can also be 
applied at $z\sim5$, where the standard lore tells us that the Universe has 
long been ionized\footnote{Note that $\Delta z=1$ at these redshifts is a 
sizable fraction ($\sim20$\%) of the Hubble time.  Therefore understanding the 
current constraints on reionization at $\zfivesix$ is important not only in 
calibrating reionization models, but also in constraining its impact on 
subsequent structure formation \citep[e.g.,][]{Busha+2010}, and in interpreting 
virtually all high redshift observations.}. Specifically, the photoionization 
rate, $\Gamma$ and mean volume-weighted neutral fraction, $\avenf$, are derived 
from $\tauGPeff$ \emph{assuming a uniform $\Gamma$} 
\citep[e.g., Fig. 7 in ][]{Fan+06}. However, reionization is expected to be 
highly inhomogeneous\footnote{Reionization driven by stellar sources is 
``inside-out'' on large-scales: ionized bubbles grow around the first, highly 
clustered sources, with their eventual coalescence (``overlap'') signaling the 
completion of reionization (e.g., \citealt*{FZH04,McQuinn+07b,TC07,Zahn+10}; 
see the recent review by \citealt{TG09}). These reionization scenarios result 
in a very inhomogenous ionization field.  If on the other hand, quasars (or 
other ionizing sources with a hard spectrum) drive the bulk of reionization, 
the ionization fields would be more uniform.  However, the abundance of quasars 
seems too low for them to contribute a significant fraction of ionizing photons 
at $z\gsim 5$ \citep{Fan+01,Jiang+08,Willott+10}; furthermore, even if faint 
quasars missed by existing surveys were present in sufficient abundance, they 
would overproduce the unresolved soft X-ray background \citep*{DHL04}.  
Therefore, even in extreme scenarios involving substantial pre-ionization by 
quasars \citep[e.g.,][]{VG09}, the late stages of reionization are expected to 
be inhomogeneous.}, with pre-overlap neutral regions having a very weak 
ionizing background (and hence low $\Gamma$).  Therefore, the {\it a priori} 
assumption of a uniform $\Gamma$ is tantamount to assuming that reionization 
has already completed; using the derived $\Gamma$ or $\avenf$ as evidence of 
this is circular logic\footnote{Despite the sizable spatial fluctuations in 
$\Gamma$ post-reionization, averaged statistics such as $\tauGPeff$ are still 
dominated by the spatial fluctuations in the density field.  Therefore the 
assumption of a uniform $\Gamma$ is valid {\it post-reionization}, since it 
only biases the inferred value of $\Gamma$ by a few percent \citep{MF09}.}.  
The Lyman forests can be made opaque by both (i) regions with trace amounts of 
neutral hydrogen inside the ionized IGM ($x_{\rm HI}\sim10^{-5}$--$10^{-4}$), 
and (ii) the neutral patches pre-overlap ($x_{\rm HI}\sim$0.1--1).  
Observationally, we cannot distinguish between these.  Theoretically, 
uncertainties in the relevant parameters are large enough to allow a sizable 
contribution from (ii), i.e. $\avenf\lsim$ tens of percent.

Given these complications, how best can existing quasar spectra be used to 
constrain reionization (i.e. $\avenf$), in the least model-dependent fashion? 
The strength of the Ly$\alpha$ and Ly$\beta$ transitions is such that even 
trace amounts of neutral hydrogen will suppress the observed flux well below 
the detection limits of current instruments, thus any remaining neutral patches 
along the lines-of-sight to distant quasars will generate ``dark'' pixels in 
the spectra of those quasars. \citet{M10} argued that the simple metric of the 
covering fraction of these dark spectral patches provides the least 
model-dependent constraint on the neutral fraction. This constraint is an upper 
limit since it does not differentiate between the above-mentioned (i) and (ii).
At $z\gsim5$, the contributions to absorption from the thickening Lyman forests 
and from the evaporating residual \HI\ patches will be exceedingly difficult to 
disentangle without detailed modeling of the IGM.  {\it Even with complete 
model confidence}, the current observational sample is insufficient to 
constrain $\avenf$ to below the percent level \citep{M10}.

In this work, we ignore these complications and instead {\it present a simple 
upper limit on the neutral fraction at $\zfivesix$ using the covering fraction 
of dark pixels}. While this limit is highly conservative since it does not 
``model-out'' the substantial absorption from residual \HI\ in the ionized IGM, 
we stress that this is the only observational constraint derived from existing 
quasar spectra free of a priori, model-dependent assumptions.  As such it is a 
natural first step towards interpreting the observations.

This work is organized as follows.  In \S \ref{sec:data}, we describe our 
observational sample of $z\sim6$ quasar spectra, while in \S\ref{sec:dark} we 
discuss our methodology for deriving the covering fraction of dark pixels from 
this sample. In \S \ref{sec:example}, we show example spectra. In 
\S \ref{sec:results} we present the corresponding upper limits 
on $\avenf$, which are the main result of this work.  In \S\ref{sec:LOS}, we 
make simple estimates of the sightline-to-sightline variance during 
reionization, and in \S \ref{sec:future} we discuss future work.  Finally, we 
summarize our results in \S \ref{sec:conc}.  We quote all quantities in 
comoving units using a standard $\Lambda$CDM cosmology with parameters 
($\Omega_\Lambda$, $\Omega_{\rm M}$, $\Omega_b$, $n$, $\sigma_8$, $H_0$) 
= (0.72, 0.28, 0.046, 0.96, 0.82, 70 km s$^{-1}$ Mpc$^{-1}$),
consistent with recent results from the {\it WMAP} satellite
\citep{Dunkley+09,Komatsu+09}. 

\begin{table}
 \begin{center}
 \caption{Keck ESI spectra\label{tab:speclist}}
 \begin{tabular}{lrrrrc}
 \hline
 Object & $z$ & $z_{\rm AB}$ & $t_{\rm exp}$ (hr) & $\tau_{\rm GP,lim}^\alpha$ & Ref. \\
 \hline
J0927+2001 &    5.77 &   19.88 &    0.33 &    3.2 & 1,8 \\
J0836+0054 &    5.81 &   18.74 &    0.33 &    4.0 & 2 \\
J0840+5624 &    5.84 &   19.76 &    0.33 &    3.4 & 1,8 \\
J1335+3533 &    5.90 &   20.10 &    0.33 &    3.1 & 1,8 \\
J1411+1217 &    5.90 &   19.63 &    1.00 &    3.0 & 3,8 \\
J0841+2905 &    5.98 &   19.84 &    0.33 &    2.9 & 4 \\
J1306+0356 &    6.02 &   19.47 &    0.25 &    3.7 & 2,8 \\
J1137+3549 &    6.03 &   19.54 &    0.67 &    3.2 & 1,8 \\
J0353+0104 &    6.05 &   20.54 &    1.00 &    2.9 & 5 \\
J0842+1218 &    6.08 &   19.64 &    0.67 &    3.4 & 6 \\
J1623+3112 &    6.25 &   20.09 &    1.00 &    3.6 & 3,8 \\
J1030+0524 &    6.31 &   20.05 &   10.32 &    4.6 & 2,8,9 \\
J1148+5251 &    6.42 &   20.12 &   11.00 &    5.1 & 7,8,9 \\
 \hline
 \end{tabular}
 \end{center}
 References: 1) \citealt{Fan+06qsos}; 2) \citealt{Fan+01}; 3) \citealt{Fan+04}; 4) \citealt{Goto06}; 5) \citealt{Jiang+08}; 6) \citealt{deRosa}; 7) \citealt{Fan+03}; 8) \citealt{Fan+06}; 9) \citealt{White+03}.
\end{table}

\section{The data}\label{sec:data}

Our sample consists of 13 quasars at $z\sim6$ with spectroscopy from the
Echellette Spectrograph and Imager \citep[ESI;][]{esi} on the Keck II 
telescope (Table~\ref{tab:speclist}).  We use only ESI spectra in order to 
have uniform wavelength coverage and resolution, and also for the high 
signal-to-noise ratio ($S/N$) even in relatively  short exposures.  Details 
of the reduction procedures are given in \citet{White+03}.  Most of these 
spectra were included in \citet{Fan+06}; as in that work we exclude quasars 
with broad absorption line (BAL) features.  The quasar redshifts have been 
updated to match the values given in \citet{Carilli+10}, taking advantage of 
more accurate CO and \MgII\ redshifts when available.  The spectra have widely 
varying exposure times (from 15 minutes to $>10$ hours) and span 1.8 optical 
magnitudes in flux, and thus have widely differing $S/N$ levels in the Lyman 
forest regions. 

We quantify the dynamic range of each spectrum in terms of the average limiting 
optical depth in the Ly$\alpha$ and Ly$\beta$ forests. To derive these 
quantities, we fit the power-law continuum model for the rest-frame UV from 
\citet{Telfer+02} to the unabsorbed continuum redward of Ly$\alpha$. The 
average $2\sigma$ limiting optical depths in the Ly$\alpha$ forest are then 
calculated as 
$\tau_{\rm GP,lim}^\alpha = -\langle \ln(2\sigma/f_{\rm cont}) \rangle$ and 
shown in Table~\ref{tab:speclist}.  For the Ly$\beta$ forest the continuum 
flux, and hence dynamic range, is affected by absorption from the underlying 
$z\sim$4\mbox{--}5 Ly$\alpha$ forest, as the photons redshift into \lya\ 
resonance at  $1+z_\alpha = (\lambda_\beta/\lambda_\alpha)(1+z_\beta)$.  We 
account for this effect using the redshift evolution of the mean opacity given 
in \citet{Fan+06}.  On the other hand, the optical depth is reduced by the 
weaker oscillator strength ($f$) and shorter wavelength of 
Ly$\beta$~($\tau_{\rm GP} \propto f\lambda$), which increases the dynamic 
range in the Ly$\beta$ forest relative to Ly$\alpha$.
Note that the continuum fitting and the derived optical depths are strictly 
used to characterize the quality of the spectra, and do enter into the 
classification of ``dark'' pixels, which is the main result of this work 
(see \S\ref{sec:dark}).

\begin{figure}
\epsscale{1.15}
 \plotone{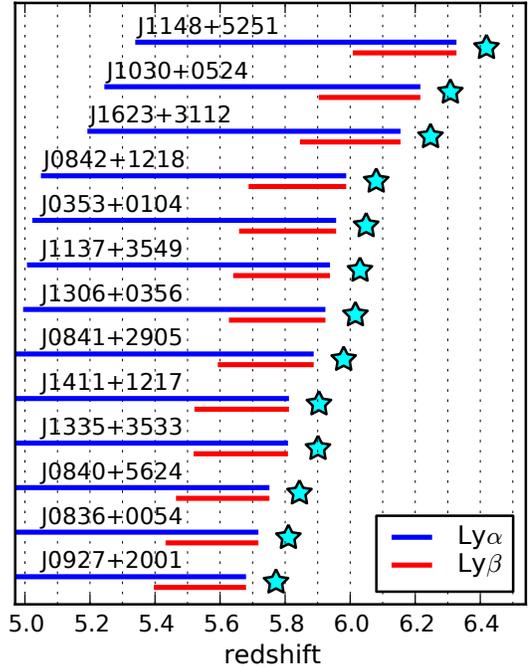}
 \caption{Ly$\alpha$/$\beta$ redshift paths spanned by the ESI spectra. 
 For each quasar the redshift is indicated by a star, the Ly$\alpha$ 
 forest by the upper (blue) line, and the Ly$\beta$ forest by the 
 lower (red) line.
 \label{fig:zpaths}}
\end{figure}

We bin the spectra into pixels of width 3.3 Mpc. The ESI spectra have a 
wavelength resolution of $R \sim 3000-6000$ \citep{Fan+06}, though they are 
sampled at a higher resolution during extraction. Binning to 3.3 Mpc results in 
pixels with $R \sim 800$ calculated from the mean of $\sim 30$ unbinned pixels.
This increases the dynamic range by $\sim20-50\%$ (as determined from the 
optical depth).\footnote{Our choice of pixel size is fairly arbitrary, and 
could be considered one of the few ``model'' uncertainties in this work. This 
scale is a factor of a few times the Jean's length in the mean-density, ionized 
IGM. Furthermore, a simple estimate of the photoevaporation timescale of 
mean-density, $\sim$3 Mpc regions exposed to the observed mean $\Gamma$ 
\citep[e.g.,][]{BH07b}, yields $\Delta z\sim 0.8$.  This suggests that even 
smaller \HI\ patches could persevere for a non-trivial time.  Neglecting such 
sub-pixel \HI\ structure likely makes our results again more conservative, 
since the associated strong opacity in the core of the Lyman line, as well as 
a potentially strong contribution from the damping wings 
\citep[e.g.,][]{Miralda-Escude1998}, would likely cause the entire 3.3 Mpc 
``pixel'' to be dark, resulting in an overestimate of the covering fraction.  
More generally. by absorbing flux in neighboring pixels, damping wing 
absorption from potential \HI\ patches would make our upper limits more 
conservative.  Unfortunately, estimating the impact of damping wings is highly 
model-dependent, requiring the distribution of \HI\ impact parameters and 
partial ionization levels.  In either case, we plan on combining our 
sky-noise-limited ESI spectra with read-noise-limited high-resolution HIRES 
spectra (e.g., \citealt{Becker+07}) in an attempt to resolve sub-structure. 
Note that we are not discussing collapsed structures 
\citep*[e.g.,][]{Iliev+05}, which are more appropriately labeled as LLSs or DLAs 
contributing to the absorption inside the ionized IGM component.}

Though the ESI spectra have been degraded in resolution in order to increase 
the dynamic range, the higher native resolution of the spectra confines the 
strong night sky emission lines to relatively small scales. We take advantage 
of this property to select bins \emph{between} the brightest sky lines. First, 
the average noise is calculated in a 3.3 Mpc window centered on each pixel in 
the unbinned spectrum.  Minima in the average noise are identified, and bins 
are placed at these minima.  This effectively identifies locations where binned 
pixels can ``fit'' between bright sky lines. Finally, the spaces between these 
bins are iteratively filled with additional bins located at the minima within 
the spaces until no more bins can fit.  We then remove from our analysis any 
pixels with too low of a threshold in optical depth, defined as a $2\sigma$ 
limit of $\tau_{\alpha} = 2.5$.

\begin{figure*}
\epsscale{2.05}
 \plotone{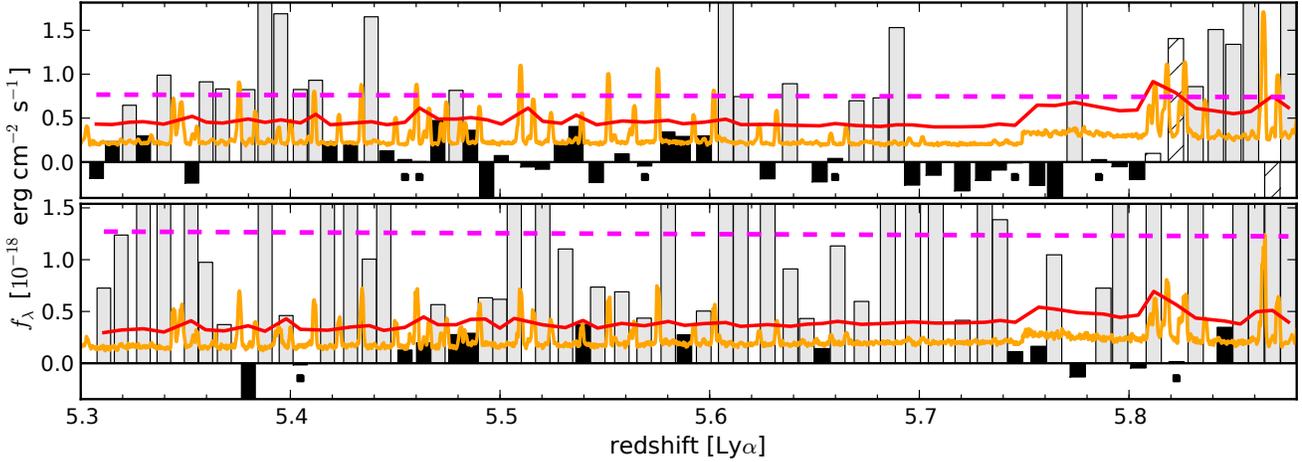}
 \caption{Ly$\alpha$ forests of J0841+2905 ($z_Q=5.98$, top panel) and
 J1306+0356 ($z_Q=6.02$, bottom panel). Pixel flux measurements are indicated 
 with bars, and shown in gray when flux is detected (according to the 
 $2\sigma$ threshold), and black for dark pixels. Dark pixels with flux 
 $\approx 0$ are indicated with small squares at negative fluxes. The smooth 
 red line shows the $2\sigma$ noise level and defines the dark threshold. 
 The orange line shows the \emph{unbinned} noise, scaled to match the 
 $1\sigma$ binned noise level. This highlights the placement of the bins 
 between the stronger sky lines. Finally, the dashed magenta line shows 
 the fitted continuum scaled by $e^{-\tau}$, with $\tau=2.5$. Pixels are 
 rejected when the binned noise exceeds this threshold, as in the case 
 with J0841+2905 for three pixels at $z>5.8$ (shown as hatched bars).
 \label{fig:forestexample}}
\end{figure*}

This process results in relatively uniform noise across the full wavelength 
range, without resorting to smoothing techniques that could amplify 
correlations between the pixels \citep[e.g.,][]{White+03}.  In addition, most 
of the binned pixels are non-adjacent, further reducing the susceptibility of 
our calculations to correlated noise introduced during extraction of the 2D 
spectra, and also to any physical correlations within the IGM on scales 
corresponding to the bin size. In the following sections we use ``pixel'' to 
refer to a 3.3 Mpc binned pixel derived from this process.

Finally, we limit the wavelength ranges of the Lyman forests used in our 
analysis in order to avoid biases and contamination.  The red edge of both the 
\lya\ and \lyb\ forests is taken to be 40 Mpc from the quasar redshift 
($\Delta z = 0.1$).  This is roughly the minimum distance required to reduce 
bias from the quasar's surrounding \HII\ morphology due to the local 
overabundance of clustered galaxies \citep{AA07,M10}. The blue edge of the 
\lya\ forest is at 
$1+z^\alpha_{\rm min} = (1+z_{\rm em})(1040 {\rm\AA}/\lambda_\alpha)$, 
chosen to avoid wavelengths affected by emission from Ly$\beta$ and \OVI\ 
\citep{Fan+06}, and contamination from the \lyb\ forest.  The blue edge of the 
Ly$\beta$ forest is at
$1+z^\beta_{\rm min} = (1+z_{\rm em})(970 {\rm\AA}/\lambda_\beta)$
in order to avoid contamination from Ly$\gamma$ emission and the Ly$\gamma$ 
forest.  We show the span of the \lya\ and \lyb\ forests in 
Fig.~\ref{fig:zpaths} for all spectra used in our analysis.

\section{Defining ``Dark'' Pixels}\label{sec:dark}

We now quantitatively define what we mean by a ``dark'' pixel.  We apply 
two criteria, as defined below, to the \lya\ and \lyb\ forests from our 
spectra.
In addition, we note that dark pixels arising from a
pre-overlap neutral region must be dark \emph{in both} \lya\ \emph{and} \lyb.
Therefore, where we have overlapping (in redshift, see Fig.~\ref{fig:zpaths}) 
spectral coverage of \lya\ and \lyb\, we also present the combined dark 
fractions.

We stress that \textit{neither method used to calculate the covering 
fraction of dark pixels directly relies on continuum fitting}, and thus we 
are relatively insensitive to the significant uncertainties associated with 
this process.

\subsection{Dark pixel fraction using a flux threshold}
\label{sec:dark_SN}

First, we define dark pixels using only the flux and noise properties of
the spectra: a pixel is dark if it has flux $f_i < n\sigma_i$, where 
$f_i$ is the measured flux and $\sigma_i$ is the measured noise of 
each pixel. Simply put, a pixel is dark if it has a flux consistent with 
zero within the given (assumed Gaussian) noise, and is not dark if it 
has $(S/N)_i>n$. We have adopted a threshold of $n=2$. For this threshold, 
$\sim2.3\%$ of dark pixels will be scattered above the threshold and 
misclassified; thus we augment our dark pixel counts and uncertainties
by this amount. 
Additionally, pixels with fluxes just above the threshold can be scattered 
below it, making our upper limits slightly more conservative (modeling this 
minor effect would require knowledge of the underlying pixel flux distribution).

Our choice of a $2\sigma$ threshold is somewhat arbitrary. We compared
results obtained with a $3\sigma$ threshold and found they are consistent --
though understandably slightly weaker -- with those presented below.

\subsection{Dark pixel fraction using negative pixels}
\label{sec:dark_negative}
 
If the noise in the extracted ESI spectrum is symmetric about zero, pixels that 
intrinsically have zero flux will have equal probabilities to be scattered 
into positive and negative values.  Therefore, we can also estimate the number 
of dark pixels by simply counting pixels with negative flux and multiplying by 
two.  Again, some pixels with small positive fluxes will be scattered to 
negative fluxes and thus incorrectly counted as dark, but this effect should 
be even smaller than when using the flux threshold method.

When we present the combined dark fraction from redshift-overlapping \lya\ 
and \lyb\ forests, we multiply the number of instances of negative flux in both 
\lya\ and \lyb\ by a factor of four.  Again, if the noise is symmetric about 
zero, then a pixel with intrinsic zero-flux will have a 1/4 chance of having 
observed negative fluxes in {\it both} \lya\ and \lyb.

Counting dark pixels using negative values relies on the assumption that the
noise is symmetrically distributed about zero for pixels with no intrinsic 
flux. Systematic effects in the reduction process, such as incorrect sky 
subtraction, could bias these values and hence our results. We do not see any
evidence for this type of bias from examination of the pixel flux distribution
\footnote{For example, examination of the flux distributions in the forest regions
normalized by the standard error shows that they are roughly Gaussian with 
$\mu=0$ and $\sigma=1$, ignoring the tail to positive flux values arising from 
flux leaks within the forest.}. 
Additionally, the deeper spectra are less susceptible to this type of bias, as 
they are created by combining many individual exposures, averaging over such 
effects.

\section{Ly$\alpha$/$\beta$ forest examples}
\label{sec:example}

Figure~\ref{fig:forestexample} shows example \lya\ forests for two objects. 
The figure demonstrates the placement of bins away from bright sky lines and 
the relatively uniform noise in the binned pixels used for the final 
calculation of the dark pixel fraction.  The forest for J0841+2905 is 
considerably darker than that of J1306+0356 in the same redshift range. This 
is likely a consequence of the lower dynamic range of the J0841+2905 
spectrum, but there could also be contributions from line-of-sight variance.  
This highlights the need for both many lines-of-sight and high quality spectra 
when deriving constraints from the Ly$\alpha$ forest.

\begin{figure}
\epsscale{1.0}
 \plotone{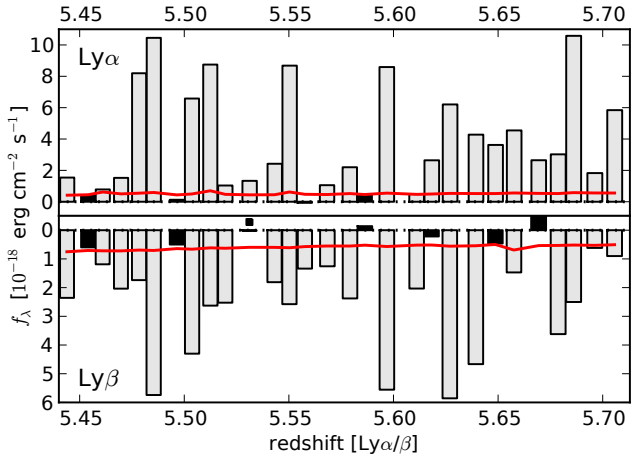}
 \caption{Comparison of the Ly$\alpha$ and Ly$\beta$ forests
 for J0836+0054 aligned in absorption redshift (note that the flux scale 
 for the Ly$\beta$ pixels is inverted for easier comparison). 
 The $2\sigma$ noise level is shown with a red line.
 The excess of dark pixels in the Ly$\beta$ forest is likely due to 
 contamination from the overlapping Ly$\alpha$ forest at 
 $1+z_\alpha = (\lambda_\beta/\lambda_\alpha)(1+z_\beta)$.
 The pixels are aligned in redshift. Comparing the two
 forests shows that only two pixels are dark in both.
 \label{fig:lyabcompare}}
\end{figure}

The Ly$\beta$ forest yields a stronger constraint on the fraction of dark 
pixels, as the relative weakness of the Ly$\beta$ transition compared to \lya\ 
results in greater dynamic range. The specific ratio of $\tau^\alpha/\tau^\beta$
varies from pixel-to-pixel depending on the sub-pixel structure of the IGM 
\citep[e.g.,][]{Songaila04,Fan+06}, and the absorption from the underlying
\lya\ forest (see \S\ref{sec:data}). Since we are treating the dark fraction 
as a direct upper limit on $\avenf$, we do not introduce any model assumptions 
about either the clumpiness of the IGM at $\zfivesix$ or the nature of the 
lower-redshift \lya\ forest into our results.  In future work, we will attempt 
to resolve and model-out the underlying \lya\ forest from the \lyb\ region, 
yielding tighter constraints, albeit at the cost of introducing model 
uncertainties (see \S\ref{sec:future}).

\begin{figure*}
\epsscale{1.0}
 \plottwo{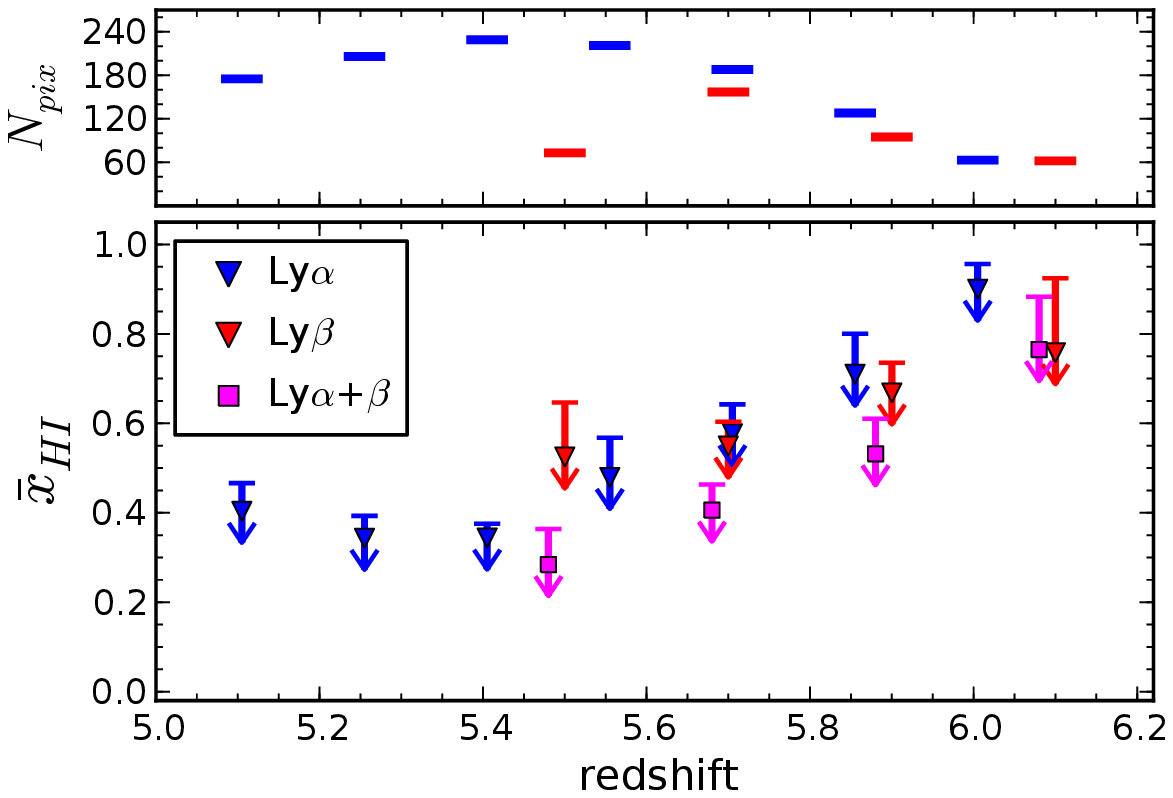}{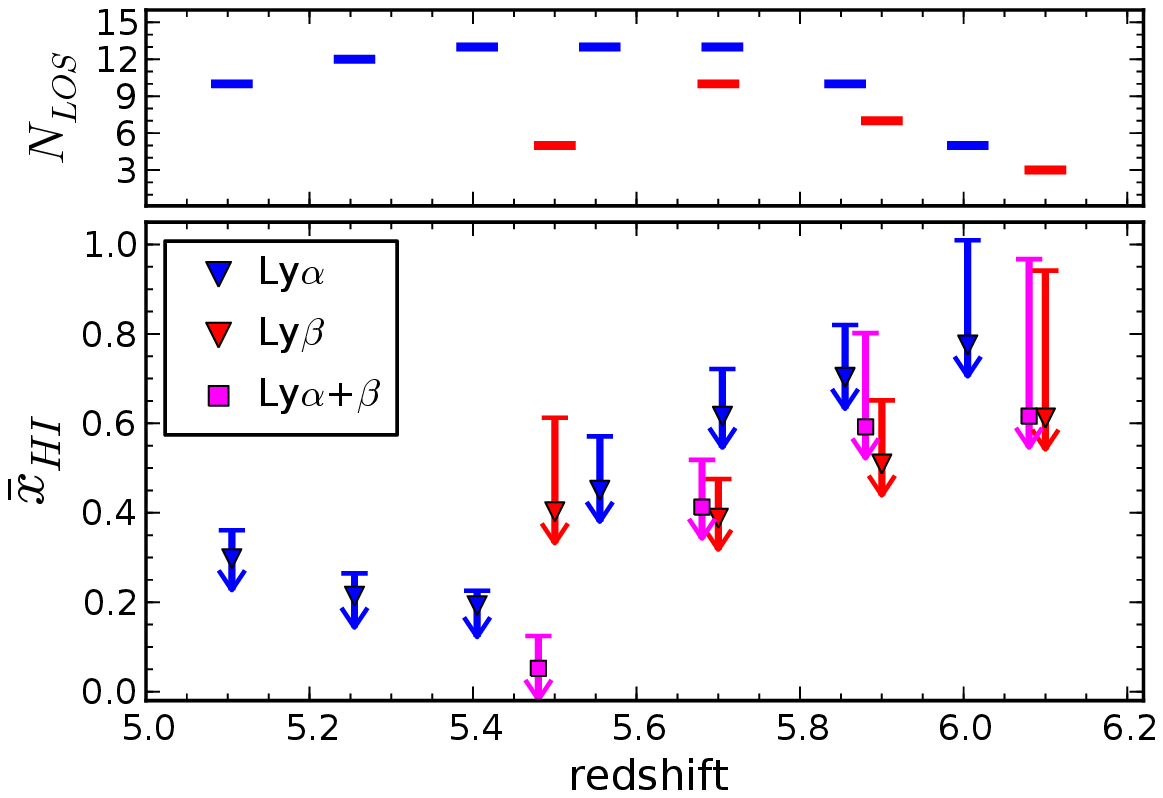}
 \caption{Upper limits on the neutral hydrogen fraction derived from Keck ESI
 spectra of $z\sim6$ quasars using Ly$\alpha$ and Ly$\beta$ forest
 absorption.
 The constraints are obtained from the covering fraction of dark pixels 
 in redshift bins of width $\Delta z = 0.15$ (Ly$\alpha$) and
 $\Delta z = 0.2$ (Ly$\beta$).
 The upper left panel shows the number of pixels in each redshift bin,
 and the upper right panel shows the number of independent lines-of-sight
 contributing to each bin.
 The lower left panel shows the constraints derived by counting dark pixels
 using a $2\sigma$ flux threshold (\S \ref{sec:dark_SN}); the lower right 
 panel shows the constraints derived from pixels with negative fluxes 
 (\S \ref{sec:dark_negative}).  
 The number of pixels and lines-of-sight are the same for both methods.
 Uncertainties are calculated from jackknife statistics (see text), as we are only
 concerned with an upper limit the uncertainty is only shown on the upper bound.
 \label{fig:xhi}}
\end{figure*}
\begin{figure*}
\epsscale{1.0}
 \plottwo{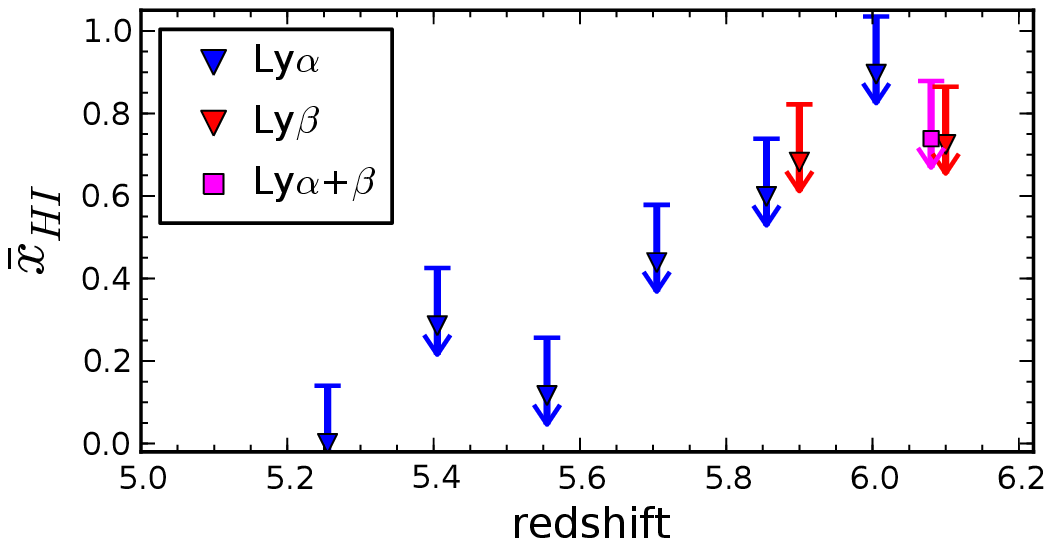}{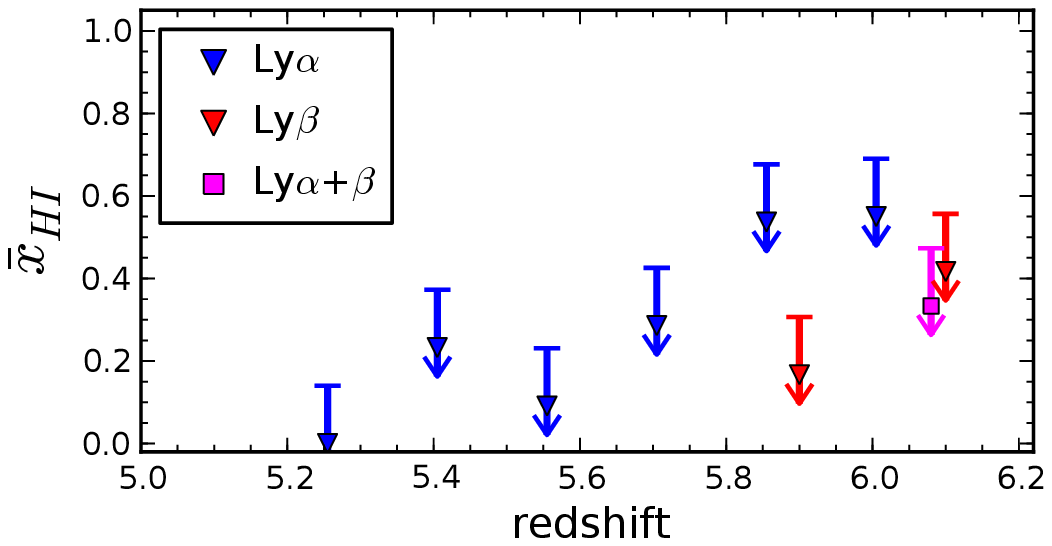}
 \caption{Same as Figure~\ref{fig:xhi}, but using only the subsample of our two 
 deepest $z>6.2$ spectra ($>10$ hour total exposure times).
 With only two spectra, these constraints are more susceptible to line-of-sight
 variance. Instead of the Jackknife errors in Fig. \ref{fig:xhi}, we plot 
 estimates of the uncertainty due to the cosmic variance of reionization at 
 $\avenf\sim0.1$ (see discussion in \S \ref{sec:LOS}).  These error bars, although 
 conservatively large, are admittedly model-dependent.
 \label{fig:xhi_deep}}
\end{figure*}

In Fig. \ref{fig:lyabcompare}, we show a comparison of the \lya\ and \lyb\ 
forests for J0836+0054, overlayed in absorption redshift.  The 2$\sigma$ noise 
level is shown with a red line, and dark pixels identified with this flux 
threshold (\S \ref{sec:dark_SN}) are shown in black.  As mentioned above, the 
contamination of the \lyb\ by the lower-redshift \lya\ forest can result in 
dark pixels even when the corresponding pixels in \lya\ are not dark.  Pixels 
corresponding to pre-overlap, highly neutral regions must be dark in both 
\lya\ and \lyb.  In this figure there are two such pixels. The intersection of 
dark pixels in both forests provides the strongest possible constraint available 
from this method; however, due to the limited dynamic range the presence of 
dark pixels in both forests is a necessary but not sufficient indication of 
neutral IGM.

Finally, it is important to remember that each of our pixels is a non-trivial 
$\sim$ 3.3 Mpc in size.  As already discussed, a single dark pixel is 
sufficiently large to be a potential tracer of pre-overlap HI, especially when 
weighted by the LOS impact parameter.  In fact, extended dark patches need not 
be a good tracer of reionization in high-redshift spectra, especially when the 
spectra begin to saturate \citep{M10}.

\section{Results}\label{sec:results}

The main results of this work are presented in Fig. \ref{fig:xhi}.  The 
covering fractions of dark pixels computed using the flux threshold 
(\S\ref{sec:dark_SN}) and negative pixel (\S \ref{sec:dark_negative}) 
statistics are directly converted into a upper limit\footnote{As mentioned 
previously, in addition to any possible pre-overlap \HI, the dark pixel 
fraction also includes contributions from \HI\ embedded in the ionized IGM 
with an optical depth greater than the available dynamic range; hence it is 
only an upper limit on $\avenf$.  Additionally, partially ionized neutral IGM 
(see discussion in \S \ref{sec:LOS}) and unresolved ionization substructure 
(on scales smaller than our pixels) make our upper limits even more 
conservative.} on the neutral hydrogen fraction and shown in the bottom left 
and right panels, respectively.  The dark fractions calculated from the \lya, 
\lyb, and combined Ly$\alpha$ + Ly$\beta$ forests are shown in blue, red, and 
magenta, respectively.  Pixels are binned in redshift bins of 
$\Delta z\sim$ 0.15 (0.2) for the \lya\ (\lyb) forest\footnote{Wide redshift 
bins decrease the sample variance, but also average over any IGM evolution.  
Our fiducial bin choice (although fairly arbitrary) is narrower than the likely 
reionization time-scales: Hubble time, photoevaporation time of pixel-sized 
regions, and the growth of the halo mass function (see, e.g., Fig.~9 in 
\citealt*{MJH06} and Fig.~1 in \citealt{Lidz+07}). Therefore, our bins are 
likely narrow enough to resolve cosmic evolution, and wide enough to have 
decent sampling of the (possible) reionization morphology in the 
$\avenf\gsim0.1$ regime (see \S \ref{sec:LOS}).}.  The upper left panel shows 
the number of pixels in each redshift bin, and the upper right panel shows the 
number of independent lines-of-sight contributing to each bin.  

The fraction of dark pixels in each bin and the associated uncertainty on
this fraction are obtained using the jackknife method. We repeat the 
calculation 13 times, each time leaving out one of the spectra. We then use
jackknife statistics to derive the mean and variance from the fractions
calculated during each resampling\footnote{Note that this is not a strict 
application of the jackknife method, as leaving out one spectrum at a time 
results in a varying number of pixels in each resampling. However, this 
variation is small.}.  Since the dark fraction is an upper limit on $\avenf$,
the derived variance is shown only in the upper limits of Figure~\ref{fig:xhi}. 
Jackknife errors may or may not fully capture the variance inherent to the 
reionization process; in any case, we estimate such cosmic variance to be 
small for our full sample (see \S \ref{sec:LOS}). Rather, the sample variance 
is most likely dominated by the disparate dynamic ranges of the spectra.

Using the flux threshold (negative pixel) method, we find upper limits of 
$\avenf \la 0.4~(0.2)$ at $5.0 \la z \la 5.5$ (fairly independent of redshift), 
increasing to $\xHI<0.7~(0.7)$ at $z\sim6$. As expected, the most powerful 
constraint for the flux threshold method is obtained from the combined \lya\ 
and \lyb\ forests, by requiring that overlapping pixels in {\it both} forests 
are dark\footnote{Note that the bins used in the analysis of the \lya, \lyb, 
and \lya\ + \lyb\ forests as shown in Fig. \ref{fig:xhi} are offset in 
redshift.  Therefore the sample variance between the bins can result in a 
\lya\ + \lyb\ dark fraction which is slightly higher than one obtained just 
from the \lyb, as is the case for the last set of points in the lower left 
panel.}.  The benefit of using the overlapping forests is not as evident for 
the negative pixel method, since it relies on events rarer by a factor of two 
(see \S\ref{sec:dark_negative}), and so has intrinsically larger sample 
variance than using just a single forest.  Including the 1-$\sigma$ jackknife 
errors degrades the $z\gsim5.5$ constraints by $\sim10\%$ ($\sim20\%$) for the 
flux threshold (negative pixel) method.

The two methods for estimating the dark covering fraction yield comparable 
results, though the negative flux method generally results in lower dark
fractions.  This could be due to many pixels with small positive flux levels 
contributing to the flux threshold method counts (\S\ref{sec:dark_SN}). By 
pushing the threshold to flux $<0$ many such pixels are eliminated, at the cost 
of higher sample variance.

There is a notable change in the evolution of the dark fraction at $z\sim 5.5$.
 This feature appears to be genuine and is due to the fact that dark gaps in 
several spectra terminate at this redshift.   Although it might be tempting to 
suggest this feature could be due to reionization, such an interpretation is 
premature.  There is no obvious, corresponding feature in the redshift 
evolution of $\tauGPeff$ \citep{Songaila04,Fan+06,Becker+07}.  Nevertheless, 
given the large scatter in the $\tauGPeff$ sample, and the difficulties in 
using $\tauGPeff$ as a reionization tracer, it will be worthwhile to 
investigate this feature with larger sample sizes and modeling.

In Fig. \ref{fig:xhi_deep}, we present results using only our two deepest 
spectra ($>$10 hour exposure times). These have an effective dynamic range of 
$\tau_{\rm GP,lim}^\alpha \sim 5$ and $\tau_{\rm GP,lim}^\beta \sim$ 7-20 
(see Table~\ref{tab:speclist}).  With just the two deepest spectra, we obtain 
mean dark fractions of $\avenf \la0.3~(0.2)$ at $z\lsim5.5$ and 
$\avenf \la 0.85~(0.5)$ at $z\sim6$, with the flux threshold (negative pixel) 
criteria.

Although these mean values are generally lower than those from our full sample  
(Fig. \ref{fig:xhi}), they are also more susceptible to cosmic variance.  
Having two LOSs increases the fractional uncertainty in the sample mean by a 
factor of $\sim2.5$ compared to our full sample of 13 LOSs, since the 
uncertainty in the sample mean scales with the number of samples as 
$1/\sqrt{\Nlos}$ (see \S\ref{sec:LOS}).  This susceptibility to cosmic variance 
depends on $\avenf$, increasing as the neutral patches become rarer.  Taking 
the specific case of $\avenf\sim0.1$ from Fig. \ref{fig:fractuncertain}, we 
see that the uncertainty in the mean dark fraction (due to reionization) from 
two LOS segments of width $\Delta z =0.2$ is comparable to the mean. Reducing
this uncertainty requires more deep spectra in order to take full advantage
of the increased dynamic range. 

Even with this model-dependent, though conservative, choice of cosmic variance 
error bars, our deepest two spectra yield $\avenf\lsim 0.3$ at 
$5 \lsim z \lsim 5.5$, comparable with those obtained from the full sample. 
Furthermore, using the negative pixel criteria in the right panel of 
Fig.~\ref{fig:xhi_deep}, we are able to place a robust upper limit at 
$z\sim6.1$ of $\avenf\lsim 0.5$.

Several model-dependent constraints on reionization at $z\sim6$ have been 
derived from other astrophysical probes such as: (1) the size of the proximity 
zone around quasars 
(\citealt*{WLC05,Fan+06,Carilli+10}, but see \citealt*{MHC04,BH07a,Maselli+07});
 (2) a claimed detection of damping wing absorption from neutral IGM in quasar 
spectra 
(\citealt{MH04,MH07}, but see \citealt{MF08});
 (3) the {\it non}-detection of intergalactic damping wing absorption in a 
gamma ray burst spectrum 
(\citealt{Totani+06}, but see \citealt{McQuinn+08});
 and (4) the number density and clustering of Ly$\alpha$ emitters 
(\citealt*{MR04,HC05,FZH06,Kashikawa+06,McQuinn+07a}, but see
\citealt*{DWH07,MF08b,Iliev+08}, Dijkstra, Mesinger, \& Wyithe, in preparation).
  All of these constraints are controversial, with considerable uncertainties.  
Our limit of $\avenf(z=6.1)\lsim0.5$ is  comparable to or better than 
these existing upper limits, and is much more robust (the model-dependence 
comes in the form of the cosmic variance error bar, and our choice here is 
conservative).

At lower redshifts ($\zfivesix$), previous claims of a highly-ionized IGM 
($\avenf\sim10^{-5}$--$10^{-4}$; e.g., Fig. 7 in \citealt{Fan+06}) are derived 
under the a priori assumption of a uniform background.  Given that reionization 
by stellar sources is highly inhomogenous, this carries the implicit assumption
that reionization is over, and therefore the standard 
$\tauGPeff \rightarrow \avenf$ conversion cannot be used to place direct 
constraints on reionization, even with complete confidence in the density 
distribution of the IGM. The upper limits we present in Fig.~\ref{fig:xhi} are 
{\it the first direct, model-independent constraints on $\avenf$ at these 
redshifts}.

\section{Cosmic Variance During Reionization}
\label{sec:LOS}

The jackknife error bars in Fig. \ref{fig:xhi} show the sightline-to-sightline 
standard deviation in the dark pixel fraction measured from our observational 
sample. The main source of this variation is the disparate dynamic range of 
our spectra.  However, the covering fraction of dark pixels itself can 
intrinsically vary from sightline-to-sightline. Again, this dark fraction has 
a contribution from the ionized IGM and potentially the pre-overlap neutral 
IGM.  Since this paper places upper limits on $\avenf$, we concern ourselves 
with the cosmic variance of the latter, i.e., reionization.

During the final stages of reionization, the pre-overlap neutral regions 
become increasingly rare.  If the mean spacing of the neutral islands is 
comparable to or greater than the length of our LOS segments,  then our sample 
is prone to cosmic variance, and we would need many LOSs to have robust 
constraints on $\avenf$.  This intrinsic cosmic variance might not be included 
in empirically derived jackknife error bars.  To lessen the cosmic variance, 
one could extend the size of the LOS segments (i.e. redshift bins), but then 
one is averaging over redshift evolution.

Here we try to get a rough, quantitative estimate of the number of LOSs 
required for an accurate sampling of pre-overlap \HI.  We note that 
this is a highly model-dependent estimate and so a detailed treatment is not 
appropriate for this work.  We begin by noting that the covering fraction 
of neutral IGM during reionization, computed from an ensemble of LOS segments, 
can be thought of as sampling some fundamental distribution with mean 
$\mu$,\footnote{Note that since the ``neutral'' IGM during reionization could be 
partially ionized, the mean covering fraction, $\mu$, does not have to equal 
$\avenf$, but can be larger.  Because of the strength of the Lyman series transitions, 
the covering fraction statistic does not distinguish between partially ionized 
\HI\ regions, $x_{\rm HI}\gsim 0.01$, and those which are fully neutral, 
$x_{\rm HI}\sim1$.  However, the mean neutral fraction, $\avenf$, does care if 
regions are partially or fully ionized. Note that here we are only referring 
to the ``neutral'' IGM, not the residual \HI\ in the ionized IGM, which should 
be at the level of $x_{\rm HI} \lsim 10^{-4}$ \citep[e.g.,][]{BH07b}.
For example, in the fiducial model of \citet{M10}, for $\avenf=0.1$ we 
obtain $\mu=0.18$.  Therefore, we differentiate between the covering 
fraction of neutral IGM and $\avenf$.  We caution the reader that this 
difference is highly model dependent.} and standard deviation, $\sigone$.  
According to the central limit theorem, as the sample size $\Nlos$ 
increases, the sample average of the covering fraction of neutral IGM 
approaches a normal distribution with mean $\mu$ and standard deviation, 
$\sigma = \sigone/\sqrt{\Nlos}$.

\begin{figure}
\epsscale{1.02}
\plotone{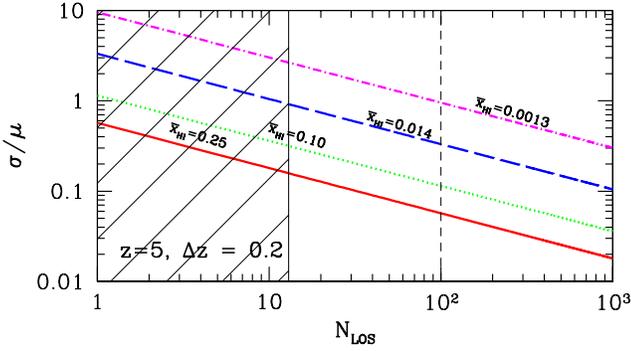}
\caption{Fractional uncertainty, $\sigma/\mu$, of the mean covering fraction of 
neutral IGM obtained from $\Nlos$ samples (see text for details).  Curves 
correspond to $\avenf=$ 0.25, 0.1, 0.014, 0.0013, bottom to top. The curves 
indicate how many $\Delta z=0.2$ LOS segments are required for their sample 
average to be accurate within the given fractional uncertainty.  The shaded 
region corresponds to the sample used in this work.  The dashed vertical line 
roughly corresponds to the current available sample of $z>5$ quasars (though 
most currently do not have deep spectra and their forests do not all overlap).  
All curves are computed at $z=5$ and assume LOS segments of length $\Delta 
z=0.2$.
\label{fig:fractuncertain}}
\end{figure}

In Fig. \ref{fig:fractuncertain}, we plot the fractional uncertainty, 
$\sigma/\mu$, of the mean covering fraction of neutral IGM obtained from 
$\Nlos$ samples.  The standard deviation of the single sample (i.e. from 1 LOS) 
distribution of the \HI\ covering fraction, $\sigone$, is taken from the 
fiducial models of \citet{M10}\footnote{These reionization morphologies are 
created with the publicly available, semi-numerical simulation DexM: \url{http://www.astro.princeton.edu/~mesinger/Sim.html}.  Simulation 
boxes are 2 Gpc on a side with 3 Mpc cells.  In the fiducial models, suites 
of reionization morphologies are created at a fixed redshift by varying the 
ionization efficiencies of atomically-cooled 
(M$_{\rm min}=2\times10^8 M_\odot$) sources, with an assumed attenuation 
length inside the ionized IGM of 50 Mpc.}.  All curves are computed at $z=5$ 
and assume LOS segments of length $\Delta z=0.2$, roughly corresponding to the 
bin size we use in this work.  Roughly speaking, the curves indicate how many 
LOS segments are required such that their average covering fraction is close 
to the true value within the given fractional uncertainty.

The shaded region in Fig. \ref{fig:fractuncertain} corresponds to the sample 
used in this work (see the top, right panel of Fig. \ref{fig:xhi}).  This means 
that at $z\lsim5.8$, where we have a sample size of $\Nlos\sim10$, we can be 
reasonably certain that our sampling of pre-overlap \HI\ is accurate to 
better than $\sim$50\% of the true value, if the true value is 
$\avenf\gsim0.1$ (i.e., a significantly neutral Universe).  Of course, if the 
Universe is more ionized, the neutral IGM patches are rarer, and thus more 
LOSs are needed to obtain the same fractional uncertainty in their sample 
mean.  For example, obtaining comparable accuracy for $\avenf\sim0.01$ 
requires $\sim100$ LOSs, which is just out of reach of the current published 
set of $z\sim5$ spectra, given that their forests do not all 
overlap\footnote{Of course constraining $\avenf$ to percent level precision 
with the Lyman forest is difficult to imagine, and so this discussion is 
mostly academic.}.  However, when the neutral patches are that rare, the dark 
covering fraction statistic is dominated by the ionized IGM, and is a very 
poor probe of reionization \citep{M10}.

Constraints derived from our two deepest spectra (Fig. \ref{fig:xhi_deep}) are 
more susceptible to cosmic variance than those derived from our full sample.  
The fractional uncertainty in the mean value of the covering fraction of 
neutral IGM is a factor of 2.5 higher for two of our spectral segments than for 
13.  From Fig. \ref{fig:fractuncertain}, we see that for two spectra at 
$\avenf=0.1$, the standard deviation of the sample average is a factor 
$\sim0.8$ times the mean value of $\mu=0.18$, i.e. $\sigma=0.8\times0.18=0.14$. 
For illustrative purposes, we use these estimates of the cosmic variance 
uncertainty for our dark fraction points in Fig. \ref{fig:xhi_deep}. Note that 
these error bars are model-dependent theoretical estimates, unlike the 
empirical jackknife errors shown for our full sample in Fig. \ref{fig:xhi},
where cosmic variance was less problematic.

The error bars in Fig. \ref{fig:xhi_deep} are appropriate for the covering 
fraction of neutral IGM, which can be substantially higher than $\avenf$, 
due to partial ionizations. In keeping with the conservative spirit of this 
work, we do not convert from the covering fraction to $\avenf$ even in the 
error bars, making them {\it overestimates} of the uncertainty on $\avenf$.

\section{Future work}\label{sec:future}

The simple upper limit on $\avenf$ obtained from this study is limited by 
contamination from absorption within the ionized IGM. This contamination can be 
addressed on two fronts: improved observations and theoretical modeling.

The observations can be improved by increasing the dynamic range of the pixels 
in order to eliminate pixels with intermediate absorption levels. 
Figure~\ref{fig:xhi_deep} shows that very deep spectra can strengthen the 
$\xHI$ constraint by a factor of two; however, obtaining deep spectra is highly 
costly in telescope time, and the payoff in dynamic range with exposure time 
rises only as $\delta\tau \propto 0.5\ln(t)$.  In addition, the Ly$\beta$ 
forest provides a stronger constraint than Ly$\alpha$, but has a much shorter 
redshift path, and thus the amount of overlapping Ly$\beta$ forest between 
quasars at a range of redshifts grows slowly with the number of objects (note 
there is only one redshift bin for 
Ly$\alpha$+Ly$\beta$ in Fig.~\ref{fig:xhi_deep}).  An intermediate approach 
of obtainining a few hours of integration time on many objects seems better: 
balancing the needs for depth to detect intermediate absorption and for many 
independent lines-of-sight to reduce cosmic variance.  This program can be 
accomplished with the sample of $\sim20$ bright ($z_{\rm AB}<20.5$) quasars at 
$z\sim6$ available today.

Aside from increasing the sample of deep spectra, there are several 
complementary observations that can improve these dark fraction constraints.  
Higher resolution spectra, for example the HIRES sample of \citet{Becker+07}, 
can be used to resolve flux leaks from dark pixels on smaller spatial scales.  
Additionally, one can use corresponding near-IR spectra (which we are in the 
process of collecting) to detect metal absorption corresponding to 
contaminating DLAs in the Ly$\alpha$/$\beta$ forests, and remove their 
contribution from the dark pixel analysis.

In addition to the observations, one could also model the expected contribution 
of the ionized IGM to the dark fraction.  Models can be calibrated using the 
latest data:  (i) analytic/numeric models of the density field 
\citep*[e.g.,][]{MHR00,TC07,BB09}; (ii) conservative (e.g., unevolving)  
extrapolations of the ionization rate from lower redshifts, where the 
ionization rate is very homogeneous and the \lya\ forest is well resolved 
\citep[e.g.,][]{Croft2004,Bolton+2006,Faucher-Giguere+2008};
and (iii) conservative extrapolations of the evolution of LLSs 
\citep*{Storrie-Lombardi+94,Stengler-Larrea+95,Peroux+03,Prochaska+10,SC10}
These approaches can be used to estimate the covering fraction of dark pixels
expected from the ionized IGM and its redshift evolution. Removing this
contamination from the total dark pixel counts leads to a model-dependent 
constraint on $\avenf$.  With sufficiently high quality data and robust limits 
on the ionized IGM contribution from models, the method presented here could 
be used to detect reionization if indeed a substantial neutral fraction 
remains at $\zfivesix$.

Furthermore, Fig. \ref{fig:xhi_deep} demonstrates the power of high dynamic 
range spectra from very deep observations within the regime where the IGM 
saturates absorption spectra.  A greater dynamic range aids in detecting flux 
from the ionized IGM (except for high column density systems such as LLSs), 
but not the neutral IGM as it is far beyond the achievable range.  Hence the 
{\it rate of change of the dark fraction as a function of the spectral dynamic 
range can provide an interesting constraint on $\avenf$}.  Comparing this rate 
of change with the slope of an assumed density PDF could yield model-dependent 
constraints on reionization.  If, for example, the dark fraction was observed 
to converge towards a constant value with increasing dynamic range (beyond 
what could be attributed to high column density systems), that would be a 
reionization signpost.  We defer more detailed analysis to future work.

\section{Conclusions}\label{sec:conc}

We present upper limits on the neutral hydrogen fraction at $\zfivesix$ 
derived from the simple, robust statistic of the covering fraction of dark
pixels in the Ly$\alpha$/$\beta$ forests of high redshift quasars. The
interpretation of quasar absorption spectra is complicated by the inhomogeneous
nature of reionization and the finite dynamic range of the Lyman forests: 
dark spectral regions can result either from residual \HI\ in the ionized 
IGM, or from pre-overlap neutral patches during the epoch of reionization.  By 
conservatively associating all dark patches with pre-overlap \HI, we provide a 
constraint that is nearly model-independent.

Using a sample of 13 $z\sim6$ quasars with Keck ESI spectra, we constrain the 
neutral fraction to be $\xHI \la 0.2$ at $z \la 5.5$, rising to $\xHI < 0.8$ 
at $z=6.1$.  We find evidence of a break in the redshift evolution of the dark 
covering fraction at $z\sim5.5$. Previous claims of a highly-ionized IGM at 
these redshifts ($\xHI\sim10^{-5}$--$10^{-4}$; e.g., Fig.~7 in 
\citealt{Fan+06}) are derived under the a priori assumption of a uniform 
background.  Given that reionization by stellar sources is highly 
inhomogeneous, this carries the implicit assumption that reionization is over, 
and therefore the standard $\tauGPeff \rightarrow \xHI$ conversion cannot be 
used to place direct constraints on reionization, even with complete confidence 
in the density distribution of the IGM.

At $z=6.1$, more stringent constraints are provided by the subsample of our 
two deepest spectra when combined with conservative, albeit model-dependent, 
estimates of cosmic variance.  Specifically, we obtain $\avenf\lsim0.5$. When 
re-evaluated in the context of an inhomogeneous reionization, the existing 
constraints on $\avenf(z\sim6)$ (derived from QSO proximity regions, damping 
wings in QSO and GRB spectra, LAE number density and clustering properties) are 
sensitive probes only of the early stages of reionization, when the Universe 
was mostly neutral.  Additionally, they are highly model-dependent.  Our 
constraint of $\avenf(z=6.1)\lsim0.5$ is comparable to or better than these 
existing constraints, and is much more robust.

We expect these limits to improve with a larger sample of deep spectra. The 
dark covering fraction statistic can also be combined with theoretical 
estimates of the absorption inside the ionized IGM to yield a model-dependent 
constraint.  Finally, the rate of change of the dark fraction as a 
function of the spectral dynamic range can provide an interesting constraint 
on $\avenf$. We will explore these possibilites in future work.

Our results show that present-day observations of Lyman forest absorption
in $z\sim6$ quasars do not rule out an end to reionization as late as
$z=5$; thus the generic statement that reionization completes by $z\sim6$
is unjustified.  Indeed, the tail end of reionization could stretch to 
$z\sim5$ without a strong observational imprint, since the final stages of 
reionization -- when the UVB became regulated by LLSs and their evolution 
\citep{FM09,Crociani+10} -- may have been extended \citep{FOh05, AA10}.

The goal of this work is not to promote a late reionization scenario.  We 
simply note that it is not ruled out by current data, and caution against 
further unjustified leaps of interpretation.  We present a direct, 
conservative upper limit on $\avenf$ that does not rely on any assumptions 
about the quasar continuum, IGM density, \HII\ morphology or ionizing 
background fields.  This can be viewed as a robust starting point for 
interpretations of high-redshift observations and theoretical models.

\vspace{+0.5cm}

\acknowledgments{
We thank George Becker and Matthew McQuinn for useful 
discussions and comments on a draft version of this paper. We acknowledge the
efforts of Bob Becker and Rick White to collect and reduce the Keck data 
presented here.
IDM and XF are supported by a Packard Fellowship for Science and Engineering 
and NSF grant AST 08-06861. 
AM is supported by NASA through Hubble Fellowship grant 
HST-HF-51245.01-A, awarded by the Space Telescope Science Institute, which is 
operated by the Association of Universities for Research in Astronomy, Inc., 
for NASA, under contract NAS 5-26555.
This work made use of the CosmoloPy package (\url{http://roban.github.com/CosmoloPy/}).\\
The authors wish to recognize and acknowledge the very significant cultural role 
and reverence that the summit of Mauna Kea has always had within the indigenous 
Hawaiian community.  We are most fortunate to have the opportunity to conduct 
observations from this mountain.
}

{\it Facilities:} \facility{Keck:II (ESI)}

\bibliographystyle{mn2e}

\end{document}